\documentclass[10pt,aps,prd,showpacs,superscriptaddress,twocolumn]{revtex4}
\usepackage{amsfonts}
\usepackage{amssymb}
\usepackage{amsmath}
\usepackage{MnSymbol}
\usepackage{graphicx}

\date{\today}

\newcommand{\pp}[1]{\left( #1 \right)}

\begin{document}

\title{Awaking the vacuum with spheroidal shells}

\author{William C. C. Lima}
\email{wccl@ift.unesp.br}
\affiliation{Instituto de F\'\i sica de S\~ao Carlos,
Universidade de S\~ao Paulo, Caixa Postal 369, 13560-970, 
S\~ao Carlos, S\~ao Paulo, Brazil}
\affiliation{Instituto de F\'\i sica Te\'orica, Universidade Estadual Paulista,
Rua Dr. Bento Teobaldo Ferraz 271, 01140-070, S\~ao Paulo, S\~ao Paulo, Brazil}

\author{Raissa F. P. Mendes}
\email{rfpm@ift.unesp.br}
\affiliation{Instituto de F\'\i sica Te\'orica, Universidade Estadual Paulista,
Rua Dr. Bento Teobaldo Ferraz 271, 01140-070, S\~ao Paulo, S\~ao Paulo, Brazil}

\author{George E. A. Matsas}
\email{matsas@ift.unesp.br}
\affiliation{Instituto de F\'\i sica Te\'orica, Universidade Estadual Paulista,
Rua Dr. Bento Teobaldo Ferraz 271, 01140-070, S\~ao Paulo, S\~ao Paulo, Brazil}

\author{Daniel A. T. Vanzella}
\email{vanzella@ifsc.usp.br}
\affiliation{Instituto de F\'\i sica de S\~ao Carlos,
Universidade de S\~ao Paulo, Caixa Postal 369, 13560-970, 
S\~ao Carlos, S\~ao Paulo, Brazil}

\begin{abstract}
It has been shown that well-behaved spacetimes may induce the vacuum 
fluctuations of some nonminimally coupled free scalar fields to go 
through a phase of exponential growth. Here, we discuss this mechanism 
in the context of spheroidal thin shells emphasizing the consequences of
deviations from spherical symmetry.
\end{abstract}


\maketitle

\section{Introduction}
\label{sec:introduction}

In a recent paper, it was shown that certain well-behaved 
spacetimes are able to induce an exponential enhancement of the vacuum 
fluctuations of some nonminimally coupled free scalar fields~\cite{lv}. 
This ``vacuum awakening mechanism'' may have consequences, in 
particular, to astrophysics, since the vacuum energy density 
of the scalar field can grow as large as the nuclear density of neutron 
stars in few milliseconds once the effect is triggered~\cite{lmv}.
As a result, the system must evolve into a new equilibrium 
configuration and eventually it should induce a burst of free scalar 
particles~\cite{llmv} (see also Ref.~\cite{novak98}-\cite{rdans12} for  
related classical analyses reaching similar conclusions). Conversely, 
the existence of classes of nonminimally coupled scalar fields can 
be unfavored by the determination of the mass-radius ratio of 
relativistic stars with known equations of state.

It is, thus, interesting to know if the main features described 
in Ref.~\cite{lmv} are preserved when assumptions as staticity 
and spherical symmetry are relaxed. In this paper, we investigate the 
vacuum awakening mechanism in the context of thin static spheroidal 
shells.  This will allow us to explore the consequences of deviations 
from sphericity, while avoiding complications concerning 
uncertainties about the interior spacetime of nonspherical 
compact sources.

The paper is organized as follows. In Sec.~\ref{sec:spacetime}, 
we follow Ref.~\cite{mccrea} and present the general properties 
of the shell spacetime, emphasizing the assumptions which were 
made in order to obtain the particular class of solutions that
we investigate. In Sec.~\ref{sec:effect}, we consider the 
quantization of a real scalar field in this background and 
proceed to discuss the vacuum awakening effect in nonspherical 
configurations. We show, in particular, that in the limit 
where spherical symmetry is recovered our results can be 
expressed in terms of known functions. In Sec.~\ref{sec:vacuum},
we discuss the exponential growth of the vacuum energy density 
in the context of spherically-symmetric shells. Section~\ref{sec:finalremarks} 
is dedicated to conclusions. We assume natural units in which $c=\hbar=G=1$ 
and metric signature $(- + + +)$ throughout the paper.

\section{Thin spheroidal shells}
\label{sec:spacetime}

Let us consider a static and axially-symmetric thin shell 
surrounded by vacuum~\cite{thinshell1}-\cite{poisson}. The most general
line element describing the external- and internal-to-the-shell 
portions of the spacetime complying with the assumptions above can be written 
as~\cite{synge}
\begin{equation}
	ds^2 = - e^{2 \lambda} dt^2 + e^{2 (\nu - \lambda)} (d\rho^2 + dz^2) 
	+ \rho^2 e^{-2\lambda} d\varphi^2 ,
\end{equation}
where $\lambda=\lambda(\rho,z)$ and $\nu=\nu(\rho,z)$ satisfy
\begin{equation}
	\frac{\partial^2 \lambda}{\partial \rho^2} + \frac{\partial^2 \lambda}{\partial z^2} 
	+ \frac{1}{\rho} \frac{\partial \lambda}{\partial \rho} = 0,
\label{lambda_eq}
\end{equation}
\begin{equation}
	\frac{\partial\nu}{\partial \rho} = \rho \left[\left(\frac{\partial \lambda}{\partial \rho} 
	\right)^2 - \left(\frac{\partial \lambda}{\partial z} \right)^2 \right],
\label{nu_eq1}
\end{equation}
and
\begin{equation}
	\frac{\partial \nu}{\partial z} = 2 \rho \frac{\partial \lambda}{\partial \rho} 
	\frac{\partial \lambda}{\partial z}.
\label{nu_eq2}
\end{equation}

The external- and internal-to-the-shell regions will 
be covered with coordinates $(t,\rho,z,\varphi)$ 
and $(\bar{t},\bar{\rho},\bar{z},\bar{\varphi})$, respectively,
where we will denote by $\cal{S}$ the 3-dimensional timelike boundary 
between them. It is worth to note that by using the spacetime 
symmetries one can choose the time and angular coordinates on 
$\cal{S}$ such that $\bar{t} =t$ and $\bar{\varphi} =\varphi$. As a result, 
we will  denote the internal coordinates simply as $(t,\bar{\rho},\bar{z},\varphi)$ 
and the shell is identified with $t={\rm const}$ sections of $\cal{S}$. Let us 
assume, moreover, that the shell lies on a $\lambda = {\rm const}$ surface:
\begin{equation}
	\lambda (\rho, z)|_{\cal{S}} = 	
	\lambda (\bar{\rho}, \bar{z})|_{\cal{S}} = 
	\lambda_0 =\textrm{const}.
\label{casca_equipotencial}	
\end{equation}
This choice leads the spacetime inside the shell to be flat with
the corresponding line element being cast as~\cite{mccrea}
\begin{equation}
	ds^2_- = - e^{2\lambda_0} dt^2 + e^{-2\lambda_0} (d\bar{\rho}^2 + d\bar{z}^2 
	+ \bar{\rho}^2d\varphi^2).
\label{ds_in0}	
\end{equation}

In order to analyze the external metric, it is convenient to perform the 
coordinate transformation $\{\rho,z\} \to \{x,y\}$ defined by
\begin{equation}
\rho \equiv a (x^2 - 1)^{1/2} (1 - y^2)^{1/2}, \;\;\; z \equiv a x y,
\end{equation}
where $x \in [1,\infty)$, $y \in [-1,1]$, and $a = {\rm const}>0$.
In terms of the $x$ and $y$ coordinates, Eq.~(\ref{lambda_eq}) reads
\begin{equation}
	\frac{\partial}{\partial x} \left[(x^2-1)\frac{\partial\lambda}{\partial x}\right] +
	\frac{\partial}{\partial y} \left[(1-y^2)\frac{\partial\lambda}{\partial y}\right] = 0.
\label{lambda_eq2}
\end{equation}
The most general solution of Eq.~(\ref{lambda_eq2}) which is regular on $y=\pm 1$
(symmetry axis) and well behaved at $x \to \infty$ (spatial infinity) can be cast as 
\begin{equation}
	\lambda = \sum_{j=0}^\infty {A_j Q_j(x) P_j(y)},\;\;\; A_j = {\rm const},
\label{general_lambda}
\end{equation}
where $P_j(z)$ and $Q_j(z)$ are the zero-order associated Legendre functions 
of first and second kinds~\cite{abramowitz}, respectively.
For the sake of 
simplicity, we will restrict ourselves to the particular class of spheroidal 
shells obtained by imposing $A_j=0$ for $j = 1, 2, \ldots$ in 
Eq.~(\ref{general_lambda}). As a result, we have
\begin{equation}
	\lambda = - \frac{\beta}{2} \ln \frac{x+1}{x-1},
	\label{lambda}
\end{equation}
where $\beta = - A_0>0$ will play the role of a geometric parameter linked to 
the shell shape. It is worthwhile to note that condition~(\ref{casca_equipotencial})
combined with the $A_1=A_2=\ldots=0$ choice restrict the possible shapes and 
stress-energy-momentum distributions of the shells considered here (see, e.g., 
Ref.~\cite{quevedo} for more general shells). Still, this class of shells is 
general enough for our purposes. 

Equation~(\ref{lambda}) implies that the shells which we consider will lie on 
$$
x = x_0 = {\rm const} >1
$$ 
surfaces. The corresponding $\nu$ solution can 
be directly obtained from  Eqs.~(\ref{nu_eq1})-(\ref{nu_eq2}):
\begin{equation}
	\nu = \frac{\beta^2}{2} \ln \frac{x^2-1}{x^2-y^2}.
\end{equation}
By combining these results, the exterior metric will read
\begin{align}
	ds^2_+ & = - \left(\frac{x - 1}{x + 1}\right)^\beta dt^2 +
	a^2 \left(\frac{x + 1}{x - 1}\right)^\beta \left(\frac{x^2 - 1}
	{x^2 - y^2}\right)^{\beta^2} 
	\nonumber \\
	&\times (x^2 - y^2) \left(\frac{d x^2}{x^2 - 1} + \frac{dy^2}{1 - y^2}\right)
	\nonumber \\
	& + a^2 \left(\frac{x + 1}{x - 1}\right)^\beta (x^2 - 1)(1 - y^2) d\varphi^2.
\label{ds_out}
\end{align}
One can see that the spacetime is asymptotically flat by taking the
$x \to + \infty$ limit in Eq.~(\ref{ds_out}). It will be shown later 
that $0<\beta<1$, $\beta=1$, and $\beta>1$ are associated with prolate, 
spherical, and oblate configurations, respectively. 

Next, we must impose continuity  of the internal and 
external induced metrics, $h_{ab}$, on $\cal{S}$. It is convenient to 
cover $\cal{S}$ with coordinates $\zeta^a=(t,y,\varphi)$, 
$a=0,2,3$, since the shell lies at $x=x_0= {\rm const}$. 
The continuity condition establishes a relationship between the internal, 
$\bar{\rho}, \bar{z},$ and external, $y$, coordinates on 
$\cal{S}$. For further convenience, however, let us replace coordinates  
$\bar{\rho}, \bar{z} $ by $\tilde{r}, \theta $ as defined below: 
$$
\bar{\rho} \equiv a \tilde{r} \sin\theta,\;\;\; 
\bar{z} \equiv a \tilde{r} \cos\theta.
$$
By doing so, Eq.~(\ref{ds_in0}) reads
\begin{align}
	ds^2_- &= - \left(\frac{x_0 - 1}{x_0 + 1}\right)^\beta dt^2
	+ a^2 \left(\frac{x_0 + 1}{x_0 - 1}\right)^\beta \nonumber \\
	& \times \left( d\tilde{r}^2 
	  + \tilde{r}^2 d\theta^2+ \tilde{r}^2 \sin^2\theta d\varphi^2\right),
\label{ds_in}
\end{align}
where we have used $\lambda_0 = -(\beta/2) \ln [(x_0+1)/(x_0-1)]$. 
Then, in order to join the metrics given by Eqs.~(\ref{ds_out}) 
and~(\ref{ds_in}) on $\cal{S}$, we impose 
$\tilde{r}|_{\cal S} = f(y)$ and $\cos\theta|_{\cal S}  = g(y)$,
where
\begin{equation}
	f (y) = \left[ \frac{(x_0^2 -1)(1 - y^2)}{1-g(y)^2} \right]^{1/2}
\label{eq_f}
\end{equation}
and
\begin{align}
	 \frac{g'(y)(1 - y^2)}{1 - g(y)^2} 
	 &= 
	  g(y)y + (1 - g(y)^2)^{1/2}
		\nonumber \\
	&\times 
	\left[\left(\frac{x_0^2 - 1}{x_0^2 - y^2}\right)^{\beta^2-1} -y^2\right]^{1/2}
\label{eq_g}
\end{align}
with $``\; ' \; " \equiv d/dy$.
The $f(y)$ function  follows immediately after $g(y)$ is determined from 
Eq.~(\ref{eq_g}) through a numerical calculation, where we fix $g(0)=0$
(to harmonize with the solution $g(y)=y$ for $\beta=1$). 
In order to guarantee that $g(y)$ is real, an extra restriction on $x_0$ must be imposed 
when $\beta > 1$:
\begin{align}
 \beta > 1 & \Rightarrow x_0 \geq \beta,
\label{x0constraint} \\  
0< \beta \leq 1 &\Rightarrow x_0>1.
 \label{x0constraint2} 
\end{align}
For $\beta=1$, Eq.~(\ref{x0constraint2}) just reflects the fact 
that the radius of a spherical shell must be larger than the Schwarzschild 
one. 

\begin{figure}[t]
\includegraphics[width=7.8cm]{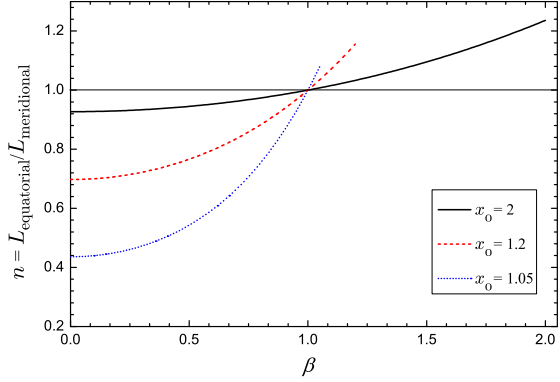}
\caption{The ratio 
$n \equiv L_{\textrm{equatorial}}/L_{\textrm{meridional}}$
is plotted as a function of $\beta$ for shells lying at 
different values of $x_0 = {\rm const}$. For $0<\beta<1$, 
$\beta = 1$, and $\beta>1$ we have prolate ($0<n< 1$), 
spherical ($n=1$), and oblate ($n > 1$) shells, respectively. 
The maximum  value of $n$, $n_{\rm max} \approx 1.3$,  
corresponds to an oblate configuration lying at 
$x=x_0=\beta \to \infty$ for which the equatorial diameter 
is about twice as large as the polar one. 
Shells with $n \to 0$ correspond to infinitely thin 
and long rods, $0<\beta<1$, lying at $x=x_0 \to 1$.}
\label{fig:form}
\end{figure}

In order to investigate the dependence of the shell shape on 
$\beta$ and $x_0$, we calculate 
\begin{align}
n & \equiv L_{\textrm{equatorial}}/L_{\textrm{meridional}}
\nonumber \\
& = \frac{(1 - 1/x_0^2)^{(1-\beta^2 )/2}} {{}_2F_1 (1/2,(\beta^2 - 1)/2,1;x_0^{-2})},
\label{n_def}
\end{align}
where  $L_{\textrm{meridional}}$  and $L_{\textrm{equatorial}}$  are the 
meridional ($\varphi = \textrm{const}$) and equatorial ($y = 0$) proper 
lengths, respectively, taken on some $t={\rm const}$ hypersurface and
${}_2F_1(a,b,c;z)$ is the hypergeometric function.  We note that 
the shell will be prolate ($0<n<1$), spherical ($n=1$), and   oblate ($n>1$)
for $0<\beta < 1$, $\beta=1$, and $\beta >1$, respectively 
(see Fig.~\ref{fig:form}). The maximum oblateness (associated with the maximum $n$ 
value) can be obtained from Eq.~(\ref{n_def}) 
combined with Eq.~(\ref{x0constraint}):   
\begin{equation}
	n_{\textrm{max}} \equiv \lim_{x_0=\beta\to\infty}{n} = {e^{1/4}}/{I_0(1/4)} 
	\approx 1.3,
\end{equation}
where $I_\nu(z)$ is the modified Bessel function of first kind.
This limit should not be viewed as a general restriction to arbitrary 
oblate shells but rather a consequence of the assumptions discussed
below Eq.~(\ref{lambda}).  There is no 
similar restriction for prolate configurations, since $n$ may take 
arbitrarily small (positive) values. For the sake of further convenience,
it is also useful to calculate, at this point, the shell proper area 
as a function of $\beta$ and $x_0$: 
\begin{align}
	A &= 4 \pi x_0^2 a^2 \left(\frac{x_0 + 1}{x_0 - 1}\right)^\beta
	\left(1-\frac{1}{x_0^2}\right)^{(\beta^2 + 1)/2}
	 \nonumber \\
	&\times {}_2F_1 \left( \frac{1}{2},\frac{\beta^2 - 1}{2}, \frac{3}{2}; \frac{1}{x_0^{2}} \right).
\label{area}
\end{align}

By establishing the interior and exterior metrics, the shell 
stress-energy-momentum tensor is also fixed:
\begin{equation}
	T^{\mu\nu} = S^{ab} e^\mu_a e^\nu_b \delta(\ell),
\label{Tmunu}	
\end{equation}
where $\ell$ is the proper distance along geodesics intercepting 
orthogonally $\cal{S}$ (such that $\ell<0$, $\ell=0$, and $\ell>0$  
inside, on, and outside $\cal{S}$, respectively), 
$e_a^\mu \equiv \partial x^\mu/\partial \zeta^a$ are the components 
of the coordinate vectors $\partial/\partial \zeta^a$ defined on 
$\cal{S}$ (with $\{x^\mu\}$ being some smooth coordinate system 
covering a neighborhood of $\cal{S}$~\cite{poisson}), and 
\begin{equation}
	S^{ab} = -\frac{1}{8\pi}(\Delta K^{ab} - h^{ab} \Delta K).
\end{equation}
Here, $K_{ab}$ is the extrinsic curvature, $K \equiv K_{ab} h^{ab}$, 
and $\Delta A^{abc\ldots}_{\;\;\;mno\ldots}$ gives the discontinuity 
of $A^{abc\ldots}_{\;\;\;mno\ldots}$ across $\cal{S}$. A straightforward
calculation leads to~\cite{mccrea}
\begin{align}
	8\pi S^0_{\; 0} & = A(y) [B(y) + C(y) -2\beta (x_0^2-y^2)], 
	\label{S00} \\
	8\pi S^2_{\; 2} & = A(y) C(y), 
	\label{S22} \\
	8\pi S^3_{\; 3} & = A(y)B(y),
	\label{S33}
\end{align}
where
$$
	A(y) \equiv \frac{(x_0^2 - 1)^{-1}}{a (x_0^2 - y^2)} \left(\frac{x_0 - 1}{x_0 + 1}\right)^{{\beta}/{2}}
	\left(\frac{x_0^2 - y^2}{x_0^2 - 1}\right)^{{(\beta^2 - 1)}/{2}}, 
$$
\begin{align}
	B(y) &\equiv \beta^2 x_0 (1 - y^2) - U(y)^{-1} (x_0^2 - \beta^2 y^2)(x_0^2 - 1) \nonumber \\
	&+ x_0 (x_0^2 - 1), 
\nonumber
\end{align}
$$
	C(y) \equiv (x_0^2 - y^2)[x_0 - U(y)], 
$$
and
\begin{equation}
	U(y) \equiv \left(\frac{x_0^2 - 1}{1 - y^2}\right)^{1/2} \left[\left(\frac{x_0^2 - 1}{x_0^2 - 
	y^2}\right)^{\beta^2 - 1} - y^2 \right]^{1/2}.
\label{U}
\end{equation}
Then, by using Eq.~(\ref{Tmunu}) combined with Eqs.~(\ref{S00})-(\ref{S33}), we obtain that
the gravitational mass formula~\cite{wald} 
\begin{equation}
M = 2 \int_{\Sigma_t} \left( T^\mu_{\;\;\nu} - \frac{1}{2} g^\mu_{\;\;\nu} T \right)\varsigma^\nu d\Sigma_\mu
\end{equation}
yields
\begin{equation}
    M =\beta a.
\label{M}
\end{equation}
Here, $\varsigma^\nu \equiv (\partial/\partial t)^\nu$ is a global timelike Killing field,
the integral is taken on a $t={\rm const}$ Cauchy surface $\Sigma_t$, 
and $d\Sigma_\mu \equiv n_\mu d\Sigma$ with $n^\mu$ being the 
pointing-to-the-future unit vector field orthogonal to $\Sigma_t$.

\begin{figure}
\includegraphics[width=7.8cm]{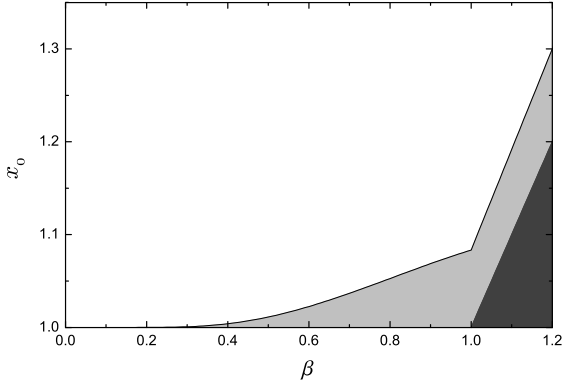}
\caption{The light gray region corresponds to values of $(\beta, x_0)$ 
which do not satisfy the dominant-energy condition, while the dark 
gray region is excluded by the constraint~(\ref{x0constraint}). The 
blank area corresponds to shell configurations satisfying the weak, 
strong and dominant energy conditions.}
\label{fig:domEC}
\end{figure}

It can be verified that the weak and strong energy conditions are
always satisfied by the stress-energy-momentum tensor~(\ref{Tmunu}). As for the dominant-energy
condition, it will be satisfied for $0 < \beta < 1$  and $\beta \geq 1$ provided 
that 
\begin{align}
	2 \beta x_0 &\geq -1 +\beta^2 + 3x_0^2 - x_0^{\beta^2}(x_0^2-1)^{1-\beta^2/2} \nonumber \\
	&-2x_0^{2-\beta^2} (x_0^2-1)^{\beta^2/2}
\label{DEC2}
\end{align}
and
\begin{equation}
	x_0 \geq ({13}/{12})\beta,
\label{DEC1}
\end{equation}
respectively (see Fig.~\ref{fig:domEC}). (One can see that for $\beta=1$ 
Eqs.~(\ref{DEC2}) and~(\ref{DEC1}) agree with each other.)
Therefore,  the matter composing this class of shells has reasonable 
physical properties for a significant range of parameters.

\section{Quantizing the field and awaking the vacuum}
\label{sec:effect}

Now, let us consider a nonminimally coupled real scalar 
field $\Phi$ with null mass defined over a spacetime of 
a spheroidal shell as discussed in Sec.~\ref{sec:spacetime}
(see Sec.~III of Ref.~\cite{llmv} for a discussion about the 
physical reasonableness of the null-mass assumption).
It will satisfy the Klein-Gordon equation
\begin{equation}
	-\nabla_\mu\nabla^\mu\Phi+\xi R\Phi=0,
\label{KG_eq}
\end{equation}
where $R$ is the scalar curvature  and 
$\xi={\rm const}$ is a dimensionless parameter. 

We follow the canonical quantization procedure and expand the 
corresponding field operator as usually:
\begin{equation}
\hat \Phi 
= 
\int d\vartheta (\eta) 
[\hat a_\eta u_\eta +\hat a_\eta^\dagger u_\eta^{*}],
\label{Phi_expandido}
\end{equation}
where $\vartheta$ is a measure defined on the set of quantum numbers $\eta$.
Here, $u_\eta$ and $u_\eta^{*}$ are 
positive and negative norm modes with respect to the Klein-Gordon inner 
product~\cite{birrell_and_davies}, respectively, satisfying 
Eq.~(\ref{KG_eq}). Then, the annihilation 
$\hat a_\eta$ and creation $\hat a_\eta^\dagger$
operators satisfy the usual commutation relations
$[\hat a_\alpha, \hat a_\beta^\dagger] = \delta (\alpha, \beta)$,
$[\hat a_\alpha, \hat a_\beta] = 0$ and the vacuum state $| 0 \rangle$
is defined by requiring $\hat a_\eta |0\rangle = 0$ for all $\eta$.    

Since the spacetime is static and axially symmetric, it is natural to look for 
positive-norm modes in the form
\begin{equation}\label{mode_form}
	u_{\eta} (t, \vec{\chi}, \varphi) = T_\sigma(t) F_{\sigma \mu} (\vec{\chi}) 
	e^{i \mu \varphi},
\end{equation}
where  $\vec{\chi} = (\tilde{r},\theta)$ and $\vec{\chi} = (x,y)$ inside and outside 
the shell, respectively, $\mu \in \mathbb{Z}$ is the azimuthal quantum number, and 
$\sigma = {\rm const}$. 
By using Eq.~(\ref{mode_form}) in Eq.~(\ref{KG_eq}), we see that $T_\sigma (t)$ obeys
\begin{equation}\label{T_eq}
	\frac{d^2}{dt^2}T_\sigma + \sigma T_\sigma = 0,
\end{equation}
while $F_{\sigma\mu} (\chi)$ satisfies
\begin{align} \label{F_in_eq}
	&-\frac{1}{a^2} \left(\frac{x_0 - 1}{x_0 + 1}\right)^{2\beta} \left[ \frac{1}{\tilde{r}^2} 
	\partial_{\tilde{r}} (\tilde{r}^2 \partial_{\tilde{r}}) + \frac{1}{\tilde{r}^2\sin\theta} 
	\partial_\theta (\sin\theta \partial_\theta ) \right. \nonumber \\
	&-\left. \frac{\mu^2}{\tilde{r}^2\sin^2\theta}\right] F^-_{\sigma\mu} = \sigma 
	F^-_{\sigma\mu}
\end{align}
and
\begin{align} \label{F_out_eq}
	&-\frac{1}{a^2}\left(\frac{x - 1}{x + 1}\right)^{2\beta}\left\{ \frac{(x^2 - y^2)^{\beta^2 - 1}}
	{(x^2 - 1)^{\beta^2}}\left\{\partial_x \left[(x^2-1)\partial_x \right] \right.\right. \nonumber \\
	& + \left.\left. \partial_y \left[(1-y^2)\partial_y \right]\right\} -\frac{\mu^2}{(x^2-1)(1-y^2)}
	\right\}F^+_{\sigma\mu} = \sigma F^+_{\sigma\mu}.
\end{align}
Here, we have assigned  labels ``$-$'' and ``$+$'' to $F_{\sigma\mu}$ in order to 
denote solutions valid inside and outside the shell, respectively. 

The solutions of Eq.~(\ref{T_eq}) will assume the following 
general forms:
\begin{equation}\label{T_osc_form}
 T_\sigma(t) \to T_\omega(t) \propto \exp (-i \omega t),\;\;\; \omega >0,
\end{equation}
for $\sigma \equiv \omega^2>0$ and
\begin{equation}\label{T_exp_form}
 T_\sigma(t) \to T_{\Omega}(t) \propto e^{\Omega t-i\pi/12}+e^{-\Omega
t+i\pi/12},\;\;\; \Omega >0,
\end{equation}
for $\sigma\equiv -\Omega^2<0$, where the latter is one of the possible 
combinations which guarantee that $u_{\sigma}$ with $\sigma<0$ 
is indeed a positive-norm mode~\cite{lv}. Equation~(\ref{T_osc_form})
is connected with the usual time-oscillating modes while
Eq.~(\ref{T_exp_form}) is associated with the so-called 
``tachyonic" modes. Tachyonic modes are responsible 
for an exponential growth of quantum fluctuations and, 
consequently, of the expectation value of the stress-energy-momentum 
tensor~\cite{lv}.  (We address to Ref.~\cite{llmv} 
for more details on the canonical quantization procedure 
in the presence of unstable modes but it is worthwhile 
to emphasize at this point that tachyonic modes do not 
violate any causality canon.) The requirement that 
these modes be normalizable determines the possible 
negative values of $\sigma$ (if any) and, thus, the existence 
(or nonexistence) of tachyonic modes. 
One sees from Eq.~(\ref{F_out_eq}) that tachyonic modes
vanish exponentially at infinity:
\begin{equation}
F_{\sigma \mu}^{+} (x,y) \to F_{\Omega \mu}^{+} (x,y) \stackrel{x \to \infty}{\propto} \exp (-\Omega ax).
\label{tachyonatinfinity}
\end{equation}

Next, let us analyze Eqs.~(\ref{F_in_eq})-(\ref{F_out_eq}) in more detail.
On account of Eq.~(\ref{Tmunu}), we have that 
$$
R = - 8\pi T = -2 \Delta K \delta (\ell).
$$ 
Then, one sees from Eq.~(\ref{KG_eq}) that $F^\pm_{\sigma\mu}$ 
should join each other continuously on $\cal{S}$: 
\begin{equation} \label{continuity}
	F^-_{\sigma\mu}(\tilde r, \theta)|_{\cal{S}}
	 	= 
	F^+_{\sigma\mu}(x,y)|_{\cal{S}},
\end{equation}
while the first derivative of $F_{\sigma\mu}$ along the direction 
orthogonal to the shell will be discontinuous: 
\begin{equation} \label{discontinuity} 
	\Delta ( dF_{\sigma\mu}/d \ell)|_{\cal{S}} 
	=  \xi \gamma(y)  F_{\sigma\mu}|_{\cal{S}}.
\end{equation}
Here, $\gamma(y) \equiv -2 \Delta K$ and we recall that 
$\Delta K = 4 \pi (S^0_{\; 0} +  S^2_{\; 2} +  S^3_{\; 3})$.
By using Eqs.~(\ref{S00})-(\ref{S33}), we obtain 
\begin{align}\label{gamma(y)}
	\gamma (y) & = - \frac{2}{a}\left(\frac{x_0 - 1}{x_0 + 1}\right)^{\frac{\beta}{2}}
	\left(\frac{x_0^2 - y^2}{x_0^2 - 1}\right)^\frac{\beta^2 - 1}{2}
	\left[ \frac{\beta^2 x_0 - \beta + x_0}{x_0^2 - 1} \right. \nonumber \\
	& + \left. \frac{x_0 (1 - \beta^2)}{x_0^2 - y^2} - \frac{1}{U(y)} 
	\left( 1 - \frac{y^2(\beta^2 - 1)}{x_0^2 - y^2}\right) - 
	\frac{U(y)}{x_0^2 - 1} \right],
\end{align}
where we recall that $U(y)$ is given by Eq.~(\ref{U}). 
For the sake of convenience, we cast Eq.~(\ref{discontinuity}) in a more explicit
form:
\begin{equation}
	\left.\pp{\frac{d x}{d \ell}\frac{\partial F^+_{\sigma\mu}}{\partial x}}\right|_{\cal{S}} 
	- \left.\pp{\frac{d \tilde{r}}{d \ell} \frac{\partial F^-_{\sigma\mu}}{\partial \tilde{r}} 
	+ \frac{d \theta}{d \ell} \frac{\partial F^-_{\sigma\mu}}{\partial \theta}}\right|_{\cal{S}} = 
	\xi \gamma(y) F_{\sigma\mu}|_{\cal{S}},
\end{equation}
where 
\begin{equation} \label{derivative_x}
	\left. \frac{d x}{d \ell}\right|_{\cal{S}} = \frac{1}{a} \left( \frac{x_0-1}{x_0+1}\right)^{\beta/2} 
	                                   \left( \frac{x_0^2-y^2}{x_0^2-1}\right)^{(\beta^2-1)/2},
\end{equation}
\begin{equation} \label{derivative_tilder}
		\left. \frac{d \tilde{r}}{d \ell}\right|_{\cal{S}} = \frac{D(y) g'(y)}{\sqrt{1-g(y)^2}},
\end{equation}
and
\begin{equation} \label{derivative_theta}
	\left.	\frac{d\theta}{d \ell}\right|_{\cal{S}} = 
	\frac{D(y) f'(y)}{f(y)^2}
\end{equation}
with
\begin{equation}
	D(y) \equiv \frac{1}{a} \left(\frac{x_0 - 1}{x_0 + 1}\right)^{\beta/2}
	\left[\frac{g'(y)^2}{1-g(y)^2} + \frac{f'(y)^2}{f(y)^2}\right]^{-1/2}.
\end{equation}

In what follows, we search for the $\xi$ parameters 
which give rise to tachyonic modes and, hence, to the 
vacuum awakening effect, once the spacetime, characterized by the values of 
$x_0$, $\beta$ and $M$, is fixed. For this purpose, we must 
look for regular $F^{\pm}_{\sigma \mu} (\vec \chi)$ functions with 
$\sigma<0$ satisfying Eqs.~(\ref{F_in_eq}) and (\ref{F_out_eq}) 
inside and outside the shell, respectively, while respecting 
Eqs.~(\ref{continuity}) and~(\ref{discontinuity}) on the shell 
and vanishing exponentially at infinity [see Eq.~(\ref{tachyonatinfinity})].

\subsection{Spherical shells}
\label{subsec:spherical}

Let us start with an analytical investigation of the conditions 
required by spherically-symmetric shells to allow the existence 
of tachyonic modes. It will be interesting in its own right and 
useful as a test for the reliability of the numerical code which 
will be used to treat more general axially-symmetric shells further.

First, we note from Eq.~(\ref{M}) that $a=M$ for $\beta=1$. 
Hence, by using the definitions $x \equiv r/M -1$  and $y \equiv \cos \theta$ 
in Eq.~(\ref{ds_out}) and $x_0 \equiv \mathsf{R}/M -1$ in Eq.~(\ref{ds_in}), 
we write the external- and internal-to-the-shell line elements as
\begin{equation}
	ds^2_+ = - (1-2M/r) dt^2 + (1-2M/r)^{-1} d r^2 + r^2 ds^2_S 
\label{ds_out_beta1}
\end{equation}
and
\begin{equation}
	ds^2_- = - (1-2M/\mathsf{R}) dt^2 
	         + (1-2M/\mathsf{R})^{-1} M^2 (d \tilde r^2 + \tilde r^2 ds^2_S),
\label{ds_in_beta1}
\end{equation}
respectively, where $ ds^2_S= d\theta^2 + \sin^2 \theta d\varphi^2$. 
In terms of the internal and external coordinates, the shell is  at
$$
\tilde r = (1-2M/ \mathsf{R})^{1/2} \mathsf{R}/M\;\; {\rm and} \;\;  r= \mathsf{R},
$$
respectively, from which we see that $\mathsf{R}$ is indeed the shell proper radius. 
In order to define a continuous radial coordinate on the shell, we introduce 
$$
r_- \equiv {M \tilde r}/{(1-2M/\mathsf{R})^{1/2}},\;\;\; r_+ \equiv r
$$
with respect to which the shell will be at $r_\pm=\mathsf{R}$.

Now, we note that the general solutions of Eqs.~(\ref{F_in_eq})-(\ref{F_out_eq}) 
can be cast in the form
\begin{equation}
	F^\pm_{\sigma\mu} (r_\pm, \theta) = \sum_{l=0}^{\infty} 
	a_{l \mu} ({\psi^\pm_{\sigma l}(r_\pm)}/{r_\pm})  P_l^\mu (\cos\theta),
\label{radial_angular_expansion}
\end{equation}
where 
$a_{l \mu} ={\rm const} $ and
$P_l^\mu(\cos \theta)$ are associated Legendre functions of the first kind, 
degree $l=0,1,2 \ldots$, and order $\mu = -l, -l+1, \ldots, l$. For the sake of 
convenience, we define the coordinates
$$
\chi_{-} \equiv r_-/(1-2M/\mathsf{R})^{1/2}, \;\;\;
\chi_{+} \equiv r_+ + 2M \ln [r_+/(2M)-1] + D,
$$
where $D={\rm const}$ is chosen such that $\chi_{-}$ and $\chi_{+}$
fit each other continuously on the shell. By using $\chi_{\pm}$, the functions 
$\psi_{\sigma l}^{\pm}$ will satisfy the ``Schr\"odinger-like" equation
\begin{equation}\label{schrodinger_type_eq}
- d^2 \psi_{\sigma l}^{\pm}/ d\chi^2_{\pm} +V_{\textrm{eff}}^{(l,\pm)}\psi_{\sigma l}^{\pm}
=\sigma\psi_{\sigma l}^{\pm}
\end{equation}
with
\begin{equation}
V_{\textrm{eff}}^{(l,-)} = (1-2M/\mathsf{R}) l(l+1)/r_-^2, 
\label{V-}
\end{equation}
\begin{equation}
V_{\textrm{eff}}^{(l,+)} = (1-2M/r_+)( l(l+1)/r_+^2 + 2M/r_+^3).
\label{V+}
\end{equation}
The discontinuity of the potential across the shell is 
$
\Delta V_{\textrm{eff}} =  2M(1-2M/\mathsf{R})/\mathsf{R}^3.
$
Although $V_{\rm eff}^{(l, \pm)}$ is positive everywhere off shell, 
the existence of tachyonic modes is still possible because the potential 
on the shell contains a delta distribution. Thus, depending on 
$\mathsf{R}/M, \xi$ and~$l$, the effective potential will be ``negative 
enough" to allow solutions of  Eq.~(\ref{schrodinger_type_eq}) for 
$\sigma<0$. This is codified in the discontinuity 
condition~(\ref{discontinuity}), which will be used later.

As a result of Eq.~(\ref{radial_angular_expansion}), the field-operator 
expansion~(\ref{Phi_expandido}) can be written in this case as
\begin{align}
\hat \Phi 
& = 
\sum_{l\mu} \int d\omega    
[\hat b_{\omega l \mu} v_{\omega l \mu} + \hat b_{\omega l \mu}^\dagger v^*_{\omega l \mu}]
\nonumber \\
&+ 
\sum_{l\mu \Omega}    
[\hat c_{\Omega l \mu} w_{\Omega l \mu} + \hat c_{\Omega l \mu}^\dagger w^*_{\Omega l \mu}],
\label{Phi_expandido_spherical}
\end{align}
where the only nonzero commutation relations between the creation and annihilation operators 
are
\begin{align}
[\hat b_{\omega l \mu}, \hat b^\dagger_{\omega' l' \mu'}] 
& =  \delta_{ll'} \delta_{\mu \mu'}\delta (\omega-\omega'),
\\
[\hat c_{\Omega l \mu}, \hat c^\dagger_{\Omega' l' \mu'}] 
& = \delta_{ll'} \delta_{\mu \mu'}\delta_{\Omega \Omega'},
\end{align}
the modes read
\begin{equation}\label{mode_form-spherical1}
	v^\pm_{\omega l \mu} = T_\omega(t) (\psi^\pm_{\omega l}(r_\pm)/r_\pm)
	Y_{l\mu} (\theta, \varphi),
\end{equation}
\begin{equation}\label{mode_form-spherical2}
  w^\pm_{\Omega l \mu} = T_\Omega(t) (\psi^\pm_{\Omega l}(r_\pm)/r_\pm)
	Y_{l\mu} (\theta, \varphi)
\end{equation}
with 
$$
Y_{l\mu} (\theta, \varphi) \equiv \sqrt{\frac{(2l+1)(l-\mu)!}{4\pi (l+\mu)!}} 
P_l^\mu (\cos\theta) e^{i \mu \varphi},
$$
and we have assigned  labels ``$\pm$'' to $v_{\omega l \mu}$ and
$w_{\Omega l \mu}$ to denote solutions valid inside and outside the shell,
following our previous notation. 

Our search for tachyonic modes equals, thus, the search for solutions 
$\psi^\pm_{\Omega l}$ of Eq.~(\ref{schrodinger_type_eq}) with 
$\sigma=-\Omega^2<0$. The regularity requirement for the normal modes 
implies that at the origin
\begin{equation}
\lim_{\chi_- \to 0} \psi^-_{\Omega l } = 0^+,
\label{psi_at_origin}
\end{equation}
where we have assumed that it approaches zero from positive values, 
since the Klein-Gordon inner product fixes the mode normalization up 
to an arbitrary multiplicative phase (which can be chosen at our 
convenience). By using Eq.~(\ref{schrodinger_type_eq}) with
Eq.~(\ref{psi_at_origin}), we conclude that  
\begin{equation}
0<\psi^-_{\Omega l }(r_-)|_{\cal S} = \psi^+_{\Omega l }(r_+)|_{\cal S},
\label{continuity_spherical}
\end{equation}
where the equality is a consequence of Eq.~(\ref{continuity}). 

On the other hand, Eq.~(\ref{discontinuity}) implies that the first 
derivative of the radial function will be discontinuous on ${\cal S}$:
\begin{align}
&
\left[\left( 1- \frac{2M}{\mathsf{R}}\right)^{1/2} \frac {d (\psi^+_{\Omega l }(r_+)/r_+)}{dr_+} 
- 
\frac{d (\psi^-_{\Omega l }(r_-)/r_-)}{dr_-}\right ]_{\cal S}
=  - 2\frac{ \xi}{\mathsf{R}}
\nonumber \\
& 
 \times \left( 1-\frac{2M}{\mathsf{R}}\right)^{-1/2} 
\left(   
2-\frac{3M}{\mathsf{R}} -2 \left( 1-\frac{2M}{\mathsf{R}}\right)^{1/2}
\right)\left. \frac{\psi^-_{\Omega l }(r_-)}{r_-} \right|_{\cal S}.
\label{discontinuity_spherical}
\end{align}
Then, $d\psi^+_{\Omega l} /d\chi_+|_{\cal S} $ will be, in general, 
a nontrivial
function of the field and shell parameters. Now, by noting from 
Eq.~(\ref{schrodinger_type_eq}) 
that 
$$
\psi^\pm_{\Omega l }\gtrless 0 \Rightarrow d^2\psi^\pm_{\Omega l }/d\chi_\pm^2 \gtrless 0,
$$ 
we conclude that either $\psi^+_{\Omega l }$ changes sign once, diverging 
negatively at infinity, or it remains always positive. Tachyonic modes 
(with $\Omega>0$) will be associated with $\psi^\pm_{\Omega l } > 0$ with 
the additional requirement that 
\begin{equation}
\lim_{\chi_+ \to +\infty} \psi^+_{\Omega l } = 0^+.
\label{tachyonic_condition}
\end{equation}
It follows, then, from Eq.~(\ref{schrodinger_type_eq}) 
that for a given shell configuration there will exist 
up to one tachyonic mode for each fixed $l$.  

In order to investigate which shell configurations give rise to 
tachyonic modes, let us first note that there always exist a negative enough 
$\sigma= -\Omega_0^2 <0$, such that 
\begin{equation}
\lim_{\chi_+ \to +\infty} \psi^+_{\Omega_0 l } = +\infty.
\label{fact}
\end{equation}
Then, if 
\begin{equation}
\lim_{\chi_+ \to +\infty} \psi^+_{0 l } = -\infty,
\label{test}
\end{equation}
it is certain that there will exist some $\Omega \in (0, \Omega_0)$ 
satisfying condition~(\ref{tachyonic_condition}). Conversely, 
if condition~(\ref{test}) is not 
verified, there will be no tachyonic mode.

\begin{figure}[t]
\includegraphics[scale=.14]{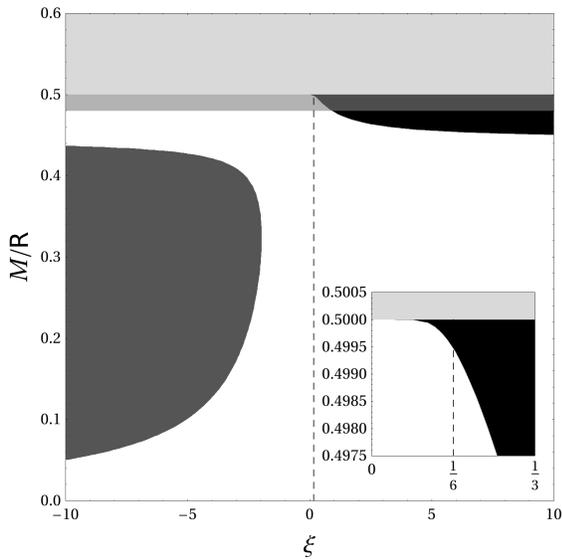}
\caption{The black and dark-gray areas depict the parameter-space region 
where $B_0/A_0<0$ leading to the ``vacuum awakening effect". 
The magnitude of the vacuum energy density on the shell grows positively and 
negatively in the black and dark-gray regions, respectively (see discussion in 
Sec.~\ref{sec:vacuum}). The light-gray strip is excluded from the 
parameter space because no static spherical shell can exist with 
$\mathsf{R} \leq 2M$, while the translucent-gray one contains those 
configurations which violate the dominant-energy condition.  The inset graph 
emphasizes that there are shell configurations which allow the presence of 
tachyonic modes for $\xi=1/6$ (vertical dashed line), although the 
dominant-energy condition is not satisfied.
}
\label{fig:spherical1}
\end{figure}

\begin{figure}[htb]
\includegraphics[scale=0.14]{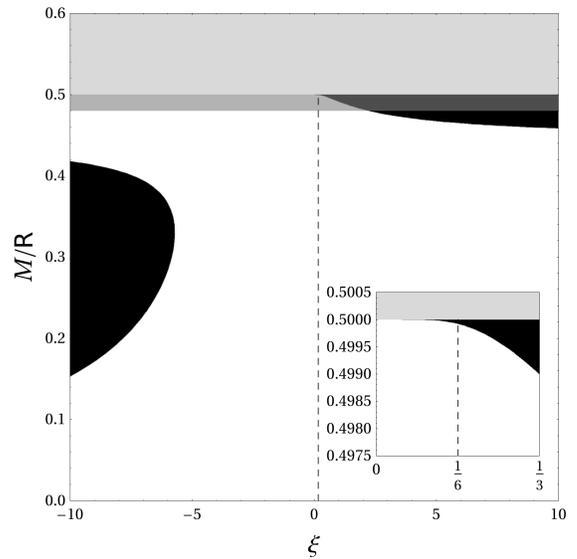}
\caption{The black areas depict the parameter-space region 
where tachyonic modes with $l=1$ are present. The light- 
and translucent-gray regions represent the same as in 
Fig.~\ref{fig:spherical1}. In contrast to the $l=0$ case, 
no analysis is performed here concerning whether the  vacuum energy 
density on the shell grows positively or negatively  
because any contribution coming from $l=1$ is dominated by the
one associated with $l=0$. 
}
\label{fig:spherical2}
\end{figure}

The solutions of Eq.~(\ref{schrodinger_type_eq}) with $\Omega =0 $ satisfying 
Eq.~(\ref{psi_at_origin}) can be written as 
\begin{equation} \label{solution_in}
	\psi^-_{0 l} ({r_-})  = A_l {r_-}^{l+1}/\mathsf{R}^l, \;\;A_l>0,
\end{equation}
\begin{equation} \label{solution_out}
	\psi^+_{0 l} (r_+) = B_l P_l (r_+/M-1)r_+ + C_l Q_l(r_+/M-1)r_+,
\end{equation}
where $A_l,B_l$ and~$C_l$ are constants. Next,
by imposing conditions~(\ref{continuity_spherical}) 
and~(\ref{discontinuity_spherical}) on the shell, we 
obtain
\begin{equation}\label{cond_0}
	\frac{B_0}{A_0} = 1- \frac{\xi\mathsf{R}}{M}
	\left(
	\frac{3M}{\mathsf{R}}-2
	+2 \sqrt{1-\frac{2M}{\mathsf{R}}} 
	\right)
	\ln\pp{1-\frac{2M}{\mathsf{R}}}
\end{equation}
and
\begin{align}\label{cond_0b}
	\frac{B_l}{A_l} = 
& - \frac{
	     [
	          (l+4\xi) ({M x_0}/{\mathsf{R}} - \sqrt{1-{2M}/{\mathsf{R}}}) 
	         -{2M\xi}/{\mathsf{R}}
	     ] 
	          Q_l(x_0)
	       }
	       {
	       ({lM}/{\mathsf{R}}) \left[
                                 P_l(x_0) Q_{l-1}(x_0) -	P_{l-1}(x_0) Q_l(x_0)
                               \right] 
	       } 
\nonumber \\
& + \frac{
	      ({lM}/{\mathsf{R}}) Q_{l-1}( x_0 )
	       }
	       {
	       ({lM}/{\mathsf{R}}) \left[
                                 P_l(x_0) Q_{l-1}(x_0) -	P_{l-1}(x_0) Q_l(x_0) 
	                           \right] 
	       } 
\end{align}
for $l=0$ and $l \geq 1$, respectively, and
\begin{equation}
\frac{C_l}{A_l} = \frac{1- (B_l/A_l) P_l(x_0)}{Q_l(x_0)}
\label{Cl/Al}
\end{equation}
for $l \geq 0$, where we recall that in the spherical case 
$x_0 = \mathsf{R}/M-1$. 
Then, by using that 
\begin{align}
&\lim_{r_+\to+\infty} Q_l(r_+/M-1) r_+ \sim r_+^{-l},
\label{Qasymptotics}
\\
&\lim_{r_+\to+\infty} P_l(r_+/M-1)r_+  \sim r_+^{l+1},
\label{Pasymptotics}
\end{align}
we see from Eq.~(\ref{solution_out}) and condition~(\ref{test}) 
that the existence of tachyonic modes with some $\Omega > 0$ requires $B_l/A_l < 0$.
($B_l/A_l=0$ corresponds to ``marginal" tachyonic modes characterized by having 
the quantum number $\Omega=0$.)  This establishes a relationship between the field 
parameter $\xi$ and the shell ratio $M/\mathsf{R}$. In Figs.~\ref{fig:spherical1}
and~\ref{fig:spherical2}, we show the parameter-space region where 
tachyonic modes with $l=0$ and $l=1$ do exist, respectively.
We note that because the smaller the $l$ the lower the 
$V_{\textrm{eff}}^{(l,\pm)}$, the existence of a tachyonic mode with 
$l=l_0$ implies the existence of tachyonic modes with $l=0, \ldots, l_0-1$. 
This can be seen in Figs.~\ref{fig:spherical1} and~\ref{fig:spherical2} 
as we note that the tachyonic-mode region for $l=1$ is contained in 
the one for $l=0$. Clearly, the existence of a single tachyonic mode is
enough to induce an exponential growth of quantum fluctuations 
leading to the vacuum awakening effect. We note, in particular, that there 
are shell configurations which allow the existence of tachyonic modes for 
the conformal field case, $\xi=1/6$. Nevertheless, it can be also seen 
from Figs.~\ref{fig:spherical1} and~\ref{fig:spherical2} that for these
configurations the dominant-energy condition~(\ref{DEC1}) is violated.

\subsection{Prolate and oblate shells}

Now, we proceed to treat the prolate ($0 < \beta < 1$) and
oblate ($\beta > 1$) spheroidal shell cases. Here, we shall 
focus our attention on the boundaries which curb the regions 
where the vacuum awakening effect occurs due to the existence 
of any tachyonic mode. These boundaries are associated with 
the presence of marginal tachyonic solutions with $\Omega =0$ 
[see discussion below  Eq.~(\ref{Pasymptotics})]. Moreover, 
following the spherically-symmetric case reasoning where the 
most likely tachyonic modes have $l=0$ (implying $\mu=0$), 
we will look for marginal tachyonic modes ($\Omega =0 
\Rightarrow \sigma=0$) with $\mu = 0$ in the axially-symmetric 
prolate and oblate cases. Then, the relevant regular solutions 
of Eqs.~(\ref{F_in_eq}) and~(\ref{F_out_eq}) which give 
rise to normalizable modes are 
\begin{equation}
	F^-_{00} = \sum_l {A_l \tilde{r}^l P_l(\cos\theta)}
\label{f_in}
\end{equation}
and
\begin{equation}
	F^+_{00} = \sum_l{ C_l Q_l(x) P_l(y)}.
\label{f_out}
\end{equation}
We note that Eqs.~(\ref{f_in})-(\ref{f_out}) generalize the
spherically-symmetric relation~(\ref{radial_angular_expansion})
with $\sigma=\mu=0$ provided that one sets $B_l = 0$ 
in Eq.~(\ref{solution_out}) [see observation within parentheses 
below Eq.~(\ref{Pasymptotics})]. 

Next, we use Eqs.~(\ref{f_in})-(\ref{f_out}) in the continuity 
condition~(\ref{continuity}) to determine the $C_l$ coefficients in 
terms of the $A_l$ ones:
\begin{equation}
	C_{l^\prime} = \frac{2l^\prime+1}{2 Q_{l^\prime}(x_0)} 
	\sum_l{A_l \int_{-1}^1{dy P_{l^\prime}(y)f(y)^l P_l[g(y)]}},
\label{Cb}
\end{equation}
where we recall that $f(y)$ and $g(y)$ are given in Eqs.~(\ref{eq_f}) 
and~(\ref{eq_g}), respectively, and we have used the orthonormality 
condition 
\begin{equation}
	\int_{-1}^1{P_n(y) P_m(y) dy} = \frac{2}{2n+1} \delta_{n m}.
\label{orth}	
\end{equation}

Once we have determined Eq.~(\ref{Cb}) for $A_{l}$ and $C_{l}$ connected 
with the marginal tachyonic modes associated with Eqs.~(\ref{f_in})-(\ref{f_out}) 
(which satisfy the proper boundary conditions), the 
frontiers which curb the unstable regions are obtained as we impose the 
first-derivative constraint~(\ref{discontinuity}). Here, it is 
convenient to note that Eq.~(\ref{discontinuity}), supplied by 
Eqs.~(\ref{f_in}), (\ref{f_out}), and~(\ref{Cb}),
can be cast as $	\sum_l{A_l G_l(y)} = 0$ for an intricate but 
otherwise known function $G_l(y)$. By expanding $G_l(y)$ in terms 
of Legendre polynomials, Eq.~(\ref{discontinuity}) can be written as 
\begin{equation} \label{sem numero}
	\sum_l{\sum_{l^\prime}{A_l k_{ll^\prime}} P_{l^\prime}(y)} = 0,
\end{equation}
where
$k_{ll^\prime} \equiv (l'+1/2)(k^{(1)}_{ll^\prime} + k^{(2)}_{ll^\prime} + k^{(3)}_{ll^\prime})$
with
\begin{equation}
	k^{(1)}_{ll^\prime} = 
	\left[\frac{1}{Q_{l^\prime}(x)}  \frac{dQ_{l^\prime}(x)}{dx}\right]_{x=x_0}
	\int_{-1}^1{dy P_{l^\prime}(y) f(y)^l P_l[g(y)]},
\end{equation}
\begin{align}
	k^{(2)}_{ll^\prime} 
	& =  -\int_{-1}^1 {dy P_{l^\prime}(y)   
  \left[ \frac{dx}{d\ell} \right]^{-1}_{\cal{S}} 
	\left\{ l f(y)^{l-1} \left[ \frac{d\tilde{r}}{d \ell} \right]_{\cal{S}} 
	P_l[g(y)]  \right.}
	\nonumber \\
	&-  \left.\frac{ l f(y)^l}{ \sqrt{1-g(y)^2}} 
	\left[ \frac{d\theta}{d \ell} \right]_{\cal{S}}  \{-g (y) P_l[g(y)] + P_{l-1}[g(y)]\} \right\}
\end{align}
and
\begin{equation}
	k^{(3)}_{ll^\prime} = -\xi \int_{-1}^1{dy P_{l^\prime}(y)  
	\left[ \frac{dx}{d\ell} \right]^{-1}_{\cal{S}} \gamma (y) 
	f(y)^l P_l[g(y)]}.
\end{equation}
Here, we recall that 
$\gamma(y)$, $d x/d\ell|_{\cal{S}}$, $d \tilde{r}/d\ell|_{\cal{S}}$, 
and $ d \theta/ d\ell|_{\cal{S}}$  are given in Eqs.~(\ref{gamma(y)}), 
(\ref{derivative_x}), (\ref{derivative_tilder}), and (\ref{derivative_theta}), 
respectively. Then, by using the orthonormality property of the Legendre 
polynomials, Eq.~(\ref{sem numero}) leads to 
\begin{equation}
	\sum_l {A_l k_{ll^\prime}} = 0.
\label{matrix}
\end{equation}
\begin{figure}[htb]
\includegraphics[width=8.5cm]{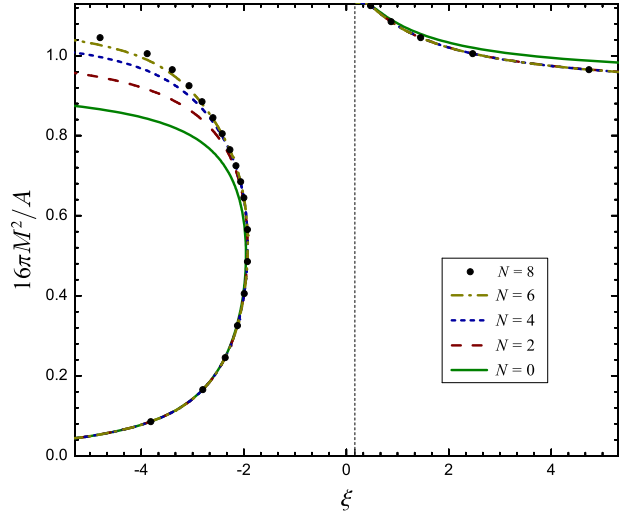}
\caption{Diagram showing how higher order $N$ approximations converge
to the actual boundaries which curb the tachyonic unstable regions
for a considerably prolate shell, $n = 0.25$. Configurations 
allowing for tachyonic modes are those to the left of the curves 
on the left-hand side and to the right of the curves on the 
right-hand side. We note that in the
$16 \pi M^2/A \ll 1$ regime, the $N=0$ approximation is already quite 
satisfactory in contrast to the  $16 \pi M^2/A \approx 1$ regime.
For $N=8$, only few points were obtained due to the computational cost.
}
\label{fig:Ns}
\end{figure}

\begin{figure}[htb]
\includegraphics[width=8.5cm]{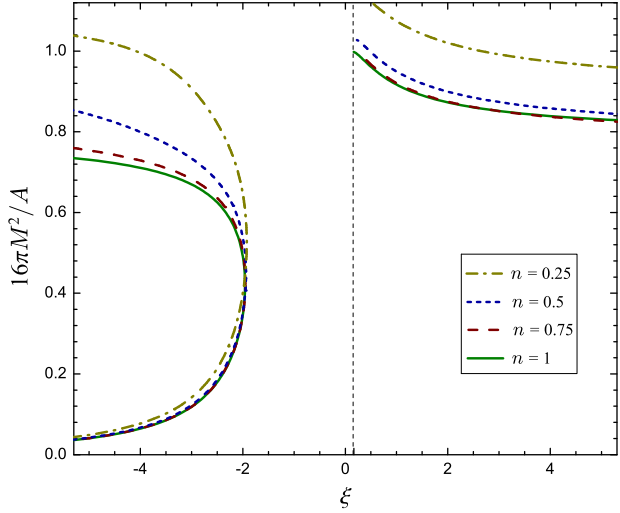}
\caption{Diagram showing the boundaries which circumscribe 
the regions where the vacuum awakening effect is triggered by 
prolate-spheroidal shells with $n = 0.25$, $0.5$ and $0.75$.
The spherical ($n=1$) case is plotted for comparison. The vertical 
dashed line indicates the conformal-coupling value $\xi = 1/6$.
Configurations allowing for tachyonic modes are those to 
the left of the curves on the left-hand side and to the right of the 
curves on the right-hand side.}
\label{fig:prolate}
\end{figure}

Let us, now, consider $k_{ll^\prime}$ as elements of a matrix $\mathcal{K}$. 
In the spherically-symmetric case, $\mathcal{K}$ is diagonal: 
$k_{ll^\prime} = g_l \delta_{ll^\prime}$ 
with $g_l$ being constants depending on the shell, $M/\mathsf{R}$, and field, 
$\xi, l$, parameters. The borderlines associated with the regions 
containing tachyonic modes for each  $l$ are obtained by 
solving $g_{l} = 0$ for  $\xi$ as a function of $M/\mathsf{R}$. 
In the absence of spherical symmetry, the corresponding borderlines 
inside which tachyonic modes exist can be obtained similarly by 
vanishing the eigenvalues of $\mathcal{K}$. The vanishing-eigenvalue 
condition can be imposed on $\mathcal{K}$ by solving the corresponding 
characteristic equation
\begin{equation}
	\det{\mathcal{K}} = 0.
	\label{det}
\end{equation}
This will drive Eq.~(\ref{matrix}) to have a nontrivial 
solution for the $A_l$ coefficients. We recall that, eventually,
all modes should be Klein-Gordon orthonormalized, which
fixes any remaining $A_l$ left free. 

For computational purposes, we truncate (the infinite matrix) $\mathcal{K}$ 
by imposing $0 \leq l,l' \leq N$ for a large enough $N$. This is justified since 
the $k_{ll^\prime}$ elements decrease as the values of $l$ or $l^\prime$ increase. 
By fixing the  $\beta$ and $x_0$ parameters, Eq.~(\ref{det}) is expected to be 
satisfied by $N + 1$ values of $\xi$, which corresponds in the spherical case to 
the fact that for a fixed $M/\mathsf{R}$ value, the boundary of the unstable
regions, associated with the marginal tachyonic modes, are at different $\xi$ 
values; each one corresponding to a distinct $l=0,\ldots ,N$ (see Figs.~\ref{fig:spherical1} 
and~\ref{fig:spherical2}). Because in the 
prolate and oblate cases we are interested in the regions where the vacuum awakening
effect occurs by the existence of any tachyonic mode, we shall 
look for the $\xi$ solutions of Eq.~(\ref{det}) which lead to the 
boundary enclosing the largest possible unstable region.
(In the spherical case, it corresponds to the boundary
of the black and dark-grey regions in Fig.~\ref{fig:spherical1} 
associated with $l=0$.) For relatively small deviations from sphericity, 
a quite reasonable approximation is already obtained by taking 
$N = 0$. By increasing $N$, we introduce higher order corrections. 
These corrections are seen to be more relevant as larger deviations 
from sphericity are considered as can be verified in Fig.~\ref{fig:Ns} 
for a considerably prolate shell.
\begin{figure}[htb]
\includegraphics[width=8.5cm]{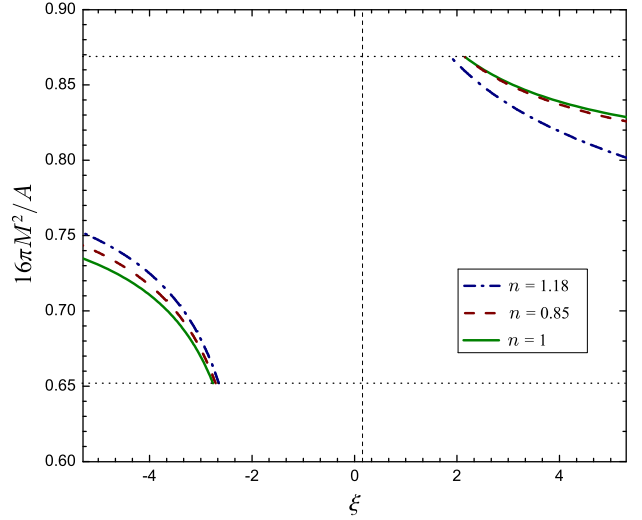}
\caption{Diagram showing the boundaries which limit the regions 
where the vacuum awakening effect is triggered by an oblate-spheroidal 
shell with $n = 1/0.85 \approx 1.18$. Prolate, $n = 0.85$, and
spherical, $n=1$, cases are plotted, as well, for the sake of comparison
(restricted to the domain of the oblate case, which is indicated by 
the horizontal dotted lines). The vertical dashed line indicates 
the conformal-coupling value $\xi = 1/6$. Configurations allowing 
for tachyonic modes are those to the left of the curves on the 
left-hand side and to the right of the curves on the right-hand 
side.}
\label{fig:oblate}
\end{figure}

In Figs.~\ref{fig:prolate} and~\ref{fig:oblate}, we show the results obtained
for some prolate and oblate shells, respectively, assuming $N = 6$.  
For the sake of clarity, we have characterized the shells by their equatorial-per-meridional 
size ratios $n$ and proper areas $A$  as given in Eqs.~(\ref{n_def}) and~(\ref{area}), 
respectively, since they have a more straightforward physical meaning than $x_0$ and $\beta$.
Figure~\ref{fig:prolate} shows that the lines which limit the regions
where the vacuum is awakened by prolate shells differ significantly from 
the spherical case for dense enough configurations, $16 \pi M^2/A \approx 1$. 
In contrast to the spherical case, where there is no equilibrium configuration
for  $16 \pi M^2/A \geq 1$, in the prolate one, $16 \pi M^2/A$ can acquire arbitrarily 
large values when $n$ is arbitrarily small. Figure~\ref{fig:oblate}
puts in context the oblate case. We recall from Sec.~\ref{sec:spacetime} that 
the degree of nonsphericity for this class of solutions is restricted on 
account of the constraint~(\ref{x0constraint}):
$\beta >1 \Rightarrow x_0 > \beta$, leading to $1 < n \lesssim 1.3$. 
This restriction reflects itself on the allowed values for 
$16 \pi M^2/A$. The excised regions at the top and bottom of Fig.~\ref{fig:oblate} 
come from this condition applied to the oblate shell considered
in the graph. From this figure, we also see that the oblate shell
with $n=1/0.85 \approx 1.18$ is more favorable to trigger the 
effect than the associated prolate one with $n=0.85$.

\section{Exponential growth of the vacuum energy density }
\label{sec:vacuum}

Finally, we investigate the exponential growth of the vacuum energy 
density  induced by the existence of tachyonic modes. Although the vacuum 
energy density will be a nontrivial point-dependent function, 
the total vacuum energy will be time conserved~\cite{lv}. Let us suppose 
that a spheroidal shell evolves from (i)~an initial static configuration 
$(x_0,\beta,M)_\textrm{in}$ where Eq.~(\ref{KG_eq}) is only allowed to 
have time-oscillating solutions to (ii)~a new  static configuration 
$(x_0,\beta,M)_\textrm{out}$ where Eq.~(\ref{KG_eq}) is permitted
to also have  tachyonic ones. Because the oscillating in-modes will 
eventually evolve into tachyonic and oscillating out-modes, the vacuum 
energy density 
$\langle \hat{T}_{00} \rangle 
\equiv \langle 0_{\rm in}| \hat{T}_{00} |0_{\rm in} \rangle$ 
will grow exponentially. Here, we assume the vacuum $|0_{\rm in}\rangle$ 
to be the no-particle state defined according to the oscillating in-modes 
(see, e.g., Ref.~\cite{llmv} for a more comprehensive discussion).  

A general expression for the exponential growth of the expectation value 
of the stress-energy-momentum tensor was calculated in Ref.~\cite{lv}. 
By applying it to the spherical shell case, we obtain the following 
leading contribution to the vacuum energy density: 
\begin{equation}\label{leading_term_erg_den0}
\langle \hat{T}_{00}\rangle
=
\langle \hat{T}^-_{00}\rangle H(-\ell)
+
\langle \hat{T}^+_{00}\rangle H(\ell)
+
\langle \hat{T}_{00}\rangle_{\cal S},
\end{equation}
where we recall that $\ell$ is the proper distance along geodesics intercepting 
orthogonally $\cal{S}$ [as defined below Eq.~(\ref{Tmunu})] and $H(\ell)$ is the 
Heaviside step function. Here, $\langle \hat{T}_{00}^- \rangle$, 
$\langle \hat{T}_{00}^+ \rangle$, and $\langle \hat{T}_{00}\rangle_{\cal S}$ 
are the vacuum contributions to the energy density inside, outside 
and on the shell, respectively, where
\begin{align}\label{leading_term_erg_den-}
\langle \hat{T}^-_{00}\rangle
&
\sim \frac{\kappa}{8\pi}
\frac{e^{2\bar{\Omega}t} }{r_-^2} \left( 1- \frac{2M}{\mathsf{R}} \right)
\frac{d}{dr_-} 
\left[ 
       \left( \frac{1-4\xi}{4\bar \Omega}  \right)  r_-^2
\right.       
\nonumber \\
&
       \left.
       \times  
       \frac {d (\psi^-_{\bar\Omega 0 }(r_-)/r_-)^2}{dr_-}
       \right],
\end{align}
\begin{align}\label{leading_term_erg_den+}
\langle \hat{T}^+_{00}\rangle
&
\sim \frac{\kappa}{8\pi}   
\frac{e^{2\bar{\Omega}t}}{r_+^2} \left( 1- \frac{2M}{r_+} \right)
\frac{d}{dr_+} 
\left[ 
       \left( \frac{1-4\xi}{4\bar \Omega}  \right)  r_+^2 
\right.       
\nonumber \\
&
       \left.
       \times 
       \left( 1- \frac{2M}{r_+} \right) 
       \frac {d (\psi^+_{\bar\Omega 0 }/r_+)^2}{dr_+}
       +\frac{\xi M}{{\bar{\Omega}}}  (\psi^+_{\bar{\Omega} 0}(r_+) /{r_+} )^2 
       \right],
\end{align}
\begin{align}\label{leading_term_erg_denS}
\langle \hat{T}_{00}\rangle_{\cal S}
&
\sim \frac{\kappa}{8\pi} e^{2\bar{\Omega}t} 
\left( 1- \frac{2M}{\mathsf{R}} \right)^{1/2}
\frac{\xi}{\mathsf{R}} \left[
              (1-4\xi) \left( 
                       \frac{3M}{\mathsf{R}}-2
                       \right.                       
\right.
\nonumber \\
&
\left.
\left.
        +2 \left(1-\frac{2M}{\mathsf{R}} \right)^{1/2} 
       \right) 
              + \frac{M}{\mathsf{R}}
\right] 
\frac{(\psi^+_{\bar{\Omega} 0}(r_+) /{r_+} )^2}{\bar{\Omega}}
\delta(\ell)
\end{align}
with $\kappa$ being a positive constant of order one related to the 
decomposition of the in-modes in terms of the out-modes and 
$\bar{\Omega}$ denoting the largest $\Omega$ selected from the set 
of all tachyonic solutions. By analyzing the factor multiplying 
the delta distribution in Eq.~(\ref{leading_term_erg_denS}),
one can verify whether the vacuum contribution to the energy density
is positive or negative on the shell. Our conclusions are depicted
in Fig.~\ref{fig:spherical1}: the black and dark-grey regions are 
associated with shell configurations where $\langle \hat{T}_{00}\rangle_{\cal S}$ 
grows positively and negatively, respectively. Similarly, one concludes 
from Eq.~(\ref{leading_term_erg_den-}) that the total vacuum energy 
inside our spherical shells is positive, null and negative when 
$\xi<1/4$, $\xi=1/4$, and $\xi>1/4$, respectively. 

As a matter of fact, eventually the scalar field and background 
spacetime must evolve into some final stable configuration, where 
tachyonic modes are not present, in order to detain  the 
exponential growth of the stress-energy-momentum tensor.  The 
precise dynamical description of how the ``vacuum falls asleep" 
again is presently under debate~\cite{pcbrs11}. In spite of the quantum 
subtleties involved in this discussion, at some point the 
scalar field is expected to loose coherence after what a 
classical general-relativistic analysis should be 
suitable. The evolution of axially-symmetric rather than 
spherically-symmetric systems may lead to new interesting 
features.

\section{Conclusions}
\label{sec:finalremarks}

It was recently shown that relativistic stars are able 
to induce an exponential enhancement of the vacuum 
fluctuations for some nonminimally coupled free scalar 
fields. In Ref.~\cite{lmv} it was assumed spherical 
symmetry to describe compact objects, which is expected 
to be a very good approximation for most relativistic 
stars~\cite{ligo}.
In this article, however, we were interested in analyzing how deviations 
from sphericity would impact on the vacuum awakening effect. 
For this purpose we have considered a class of axially-symmetric 
spheroidal shells. This has allowed us to pursue our goal, 
while avoiding concerns about how to model the interior spacetime 
of nonspherical compact sources. Fig.~\ref{fig:prolate} shows 
that for dense enough configurations, $16 \pi M^2/A \approx 1$, 
the awakening of the vacuum becomes more sensitive to prolate 
deviations from sphericity.  Fig.~\ref{fig:oblate} unveils that 
oblate shells with $n=n_0$ seem to be more efficient to awake the 
vacuum in comparison to prolate ones with $n=1/n_0$.

As a consistency check, we have performed an analytical investigation 
for the spherically-symmetric shell case in order to test the numerical 
codes used to discuss the general axially-symmetric one. It was shown, 
in particular, that in contrast to the relativistic stars analyzed
in Ref.~\cite{lmv}, spherically-symmetric shells are able to awake 
the vacuum for conformally coupled scalar fields, $\xi=1/6$ 
(see Figs.~\ref{fig:spherical1}-\ref{fig:spherical2}). The exponential 
growth of the vacuum energy density was analyzed for the 
spherically-symmetric case in Sec.~\ref{sec:vacuum}.

The present paper is part of a quest which aims at understanding the vacuum 
awakening effect in the context of physically realistic stars, where
(i)~deviations from sphericity, (ii)~rotation, and (iii)~realistic 
equations of state must be considered. In Ref.~\cite{lmv}, the authors
focused on~(iii), while in the present paper we have privileged~(i).
We are presently giving attention to~(i) and~(ii) by analyzing the vacuum 
awakening effect in the spacetime of spheroidal rotating shells~\cite{mmv}.
The full consideration of the three aspects altogether will be necessary
for a sharp prediction about what scalar fields would have their vacua 
awakened by realistic relativistic stars. It would be particularly 
interesting to see whether neutron stars would be able to awake the 
vacuum of minimally and conformally coupled scalar fields.
      
\acknowledgments

W.L. and R.M. would like to acknowledge full financial support from 
Funda\c{c}\~ao de Amparo \`a Pesquisa do Estado de  S\~ao Paulo 
(FAPESP). G.M. is grateful to FAPESP and Conselho Nacional de 
Desenvolvimento Cient\'\i fico e Tecnol\'ogico (CNPq) for partial 
suport, while D.V. acknowledges partial support from FAPESP.

\end{document}